\pdfoutput=1
\documentclass[english]{article}
\usepackage{geometry}
\usepackage{babel}
\usepackage{xcolor}
\usepackage{graphicx}
\usepackage{amsmath,amsfonts}
\usepackage{xcolor}
\usepackage{graphicx,indentfirst,subfigure,epsfig}
 \usepackage{varioref}
 \usepackage{wrapfig}
 \usepackage{subfigure}
 \usepackage{subfigmat}
 \usepackage{amsthm}
\usepackage{hyperref}
\hypersetup{colorlinks, citecolor=red, linkcolor=blue, urlcolor=red, breaklinks=true}
\usepackage{scrextend}
\usepackage{nomencl}
\makenomenclature
\usepackage{booktabs}
\usepackage{longtable}
\usepackage{soul}
\usepackage{lineno}

\usepackage[italic]{mathastext}
\usepackage{bm}
  \def\clap#1{\hbox to 0pt{\hss#1\hss}}

\providecommand{\mat}[1]{\bm{#1}}%

\providecommand{\mI}{\ensuremath{\mat{I}}}

\begin{document}

\title{Bayesian Mass Averaging in Rigs and Engines}

\author{Pranay Seshadri$^{\dagger}$\thanks{Address all correspondence to p.seshadri@imperial.ac.uk. Copyright (c) 2021 by Rolls-Royce  plc.}, Andrew B. Duncan$^\dagger$, George Thorne$^{\ddagger}$}

\date{\small{$^\dagger$Imperial College London, London, U. K. \\ $^\ddagger$Rolls-Royce plc., Derby, U. K.}}

\maketitle 

\begin{abstract}
This paper introduces the Bayesian mass average and details its computation. Owing to the complexity of flow in an engine and the limited instrumentation and the precision of the sensor apparatus used, it is difficult to rigorously calculate mass averages. Building upon related work, this paper views any thermodynamic quantity's spatial variation at an axial plane in an engine (or a rig) as a Gaussian random field. In cases where the mass flow rate is constant in the circumferential direction but can be expressed via a polynomial or spline radially, this paper presents an analytical calculation of the Bayesian mass average. In cases where the mass flow rate itself can be expressed as a Gaussian random field, a sampling procedure is presented to calculate the Bayesian mass average. Examples of the calculation of the Bayesian mass average for temperature are presented, including with a real engine case study where velocity profiles are inferred from stagnation pressure measurements.
\end{abstract}

\section{Introduction}
\label{sec:intro}
The operational state of turbomachinery is typically delineated by 1D parameter values. These include isentropic and polytropic efficiencies, pressure ratios, flow capacities, and their synthesised 1D pressures and temperatures. The averaging of three-dimensional non-uniform turbomachinery flow-fields to arrive at these 1D stagnation and static values has therefore been the subject of considerable interest and importance \cite{livesey, greitzer_tan_graf_2004}. In their seminal paper on the topic, Cumpsty and Horlock \cite{cumpsty2006averaging} rationalise that averaging should be done in line with the intended application of the average. This point was previously echoed in Pianko and Wazelt \cite{pianko1983propulsion}, who where the first to compile a comprehensive list of the thermodynamically grounded averaging methods for various flow scenarios. In what follows, we briefly introduce some of these.

For stagnation temperature averages, the mass average is appropriate as it yields the correct value of enthalpy flux as per ideal gas assumptions \cite{cumpsty2006averaging}. The mass average is given by
\begin{equation}
\bar{T}_{\dot{m}}=\frac{\int_{0}^{2\pi}\int_{r_{i}}^{r_{o}} T\left(r,\theta\right)\dot{m}\left(r,\theta\right)rdrd\theta}{\int_{0}^{2\pi}\int_{r_{i}}^{r_{o}}\dot{m}\left(r,\theta\right)rdrd\theta},
\label{equ:mass_temp}
\end{equation}
where $\dot{m} \left(r, \theta \right)$ is the mass flow rate distribution, $r_{i}$ and $r_{o}$ are the inner and outer annulus radii, and $T\left(r,\theta\right)$ indicates the stagnation temperature. The one exception to using the mass average for temperature is when averaging flow entering a choked nozzle such as the first row of high pressure nozzle guide vanes (stators) in a turbine. Here, as the flow capacity $\Gamma = \dot{m} \sqrt{c_p T} / A$ across the throat area $A$ should be preserved, the correct average is based on the square root of the temperature
\begin{equation}
\bar{T}_{\Gamma}=\left( \frac{\int_{0}^{2\pi}\int_{r_{i}}^{r_{o}} \sqrt{ T\left(r,\theta\right) }\dot{m}\left(r,\theta\right)rdrd\theta}{\int_{0}^{2\pi}\int_{r_{i}}^{r_{o}}\dot{m}\left(r,\theta\right)rdrd\theta} \right)^2,
\label{equ:cap_temp}
\end{equation}
assuming there are no variations in the specific heat capacity $c_p$ and the ratio of heat capacities $\gamma$. For stagnation pressures, the recommended average at the inlet of a turbomachine is the \emph{work average} $\bar{P}_{w}$, calculated by assuming the same power input to yield a uniform outlet pressure, under reversible and adiabatic conditions. This is given by
\begin{equation}
\bar{P}_{w} = \left( \frac{\int_{0}^{2\pi}\int_{r_{i}}^{r_{o}} T\left(r,\theta\right)\dot{m}\left(r,\theta\right)rdrd\theta}{\int_{0}^{2\pi}\int_{r_{i}}^{r_{o}} \left( T\left(r,\theta\right) / P\left(r,\theta\right)^{ ( \gamma-1 )/\gamma}  \right) \dot{m}\left(r,\theta\right)rdrd\theta} \right)^{\gamma/(\gamma - 1)},
\label{equ:press_w}
\end{equation}
where $P\left(r, \theta \right)$ is the stagnation pressure at the inlet. The second scenario where a different stagnation pressure formula is used is when considering flows channelled into a propelling nozzle, such as the flow downstream of a fan (bypass) or low pressure turbine. Here, a stagnation pressure that is based on an equivalent thrust average is used (see equation (4) in \cite{cumpsty2006averaging}). What underscores all these expressions is that they require evaluating an integral of the form $\int \int  \left(\; \cdot\; \right) \dot{m}\left(r, \theta \right) r dr d\theta$, i.e., a mass average. Thus, in scope, the ideas in this paper go beyond the determination of \eqref{equ:mass_temp}.




As long as there are no separations or asymmetries within the flow, the differences between different averaging practices are likely to be small (see Chapter 5 in Greitzer et al. \cite{greitzer_tan_graf_2004}). However, this is not always the case in the engine environment. Regardless, there is very little opportunity to adopt the aforementioned thermodynamically sound averaging practices. Instead, more often than not, the area average is widely used \cite{cumpsty2006averaging} (also see page 111 of SAE AIR1419C \cite{sae2017}). It is calculated by weighting each measurement with the cross-sectional annular area it spans, and then dividing it by the total cross-sectional area (see page 8 in Skiles 1980 \cite{skiles1980turbine}). In industry, weighting factors are used to convert this area average into other averages (see 4.3 in Seshadri et al. 2020a \cite{seshadri2020a}). However, there is no theoretical basis for this averaging practice, and no guarantee that if one were too add more sensors, one would see the area average converge to its true value (see 7.3 in \cite{seshadri2020c}). The limitations of this weighting approach have been recognised before; Pianko and Wazelt \cite{pianko1983propulsion} suggested curve fitting the (sparse) sensor data with some judgement based on the curvature of the velocity profile. Fact remains that a better assessment of the spatial distribution of the flow will likely yield a more precise area average.



Lately, there has been much interest in this singular topic with numerous authors proposing different approaches to circumferential and radial curve fitting. In Seshadri et al. 2020a and 2020b \cite{seshadri2020a, seshadri2020b}, the authors propose a multiple linear least squares model with regularisation for simultaneously fitting a Fourier series to the circumferential data and a high-degree polynomial to the radial data arising from engine temperature and pressure measurements. For estimating the unknown circumferential modes (wave numbers), they provide an algorithm which requires iterating through different frequency combinations. Their uncertainty analysis \cite{seshadri2020b} captures both the mean and variance in the resulting area average -- useful for subsequent uncertainty budgets. In \cite{lou2021reconstructing}, Lou and Key also use least squares to construct a Fourier series estimate of the circumferential variation of pressure at an isolated radial location. They pre-select certain probable circumferential modes via established guidelines based on the number of upstream and downstream vanes and struts. Practical utility of their reconstruction methodology has been demonstrated on both the static pressure field in a centrifugal compressor, and the stagnation pressure field in a three-stage axial compressor \cite{lou2}. One limitation in papers that utilise this least squares approach for reconstruction is the requirement that the number of circumferential rakes must be greater than twice the number of circumferential modes.

A problem of related interest is that of optimal rake placement. Naturally, this is closely tied to the circumferential modes present. We refer interested readers to Lou et al. \cite{lou2021reconstructing} and Chilla et al. \cite{chilla2020reducing} for such strategies, but omit further discussion on this topic here.

A distinct approach to spatial field approximation is presented in Seshadri et al. 2020c \cite{seshadri2020c}, where the authors introduce a Bayesian methodology for simultaneously interpolating along the radial and circumferential directions. The resulting profile is a Gaussian random field, defined across an annulus. The authors jointly model the circumferential and radial variations using a product kernel comprised of Fourier basis and squared exponential terms. In instances where no information regarding the circumferential modes is provided, they utilise a Bayesian analogue of compressed sensing (see \cite{piironen2017sparsity}) to recover potential candidate harmonics. Overall, their approach is robust even in the presence of missing or anomalous measurements. Additionally, through their methodology, the spatial area average---termed the Bayesian area average---is obtained in closed form. The Bayesian area average facilitates rigorous area average values when both (i) the sensor measurements have uncertainties, and (ii) there is a dearth of instrumentation.

This paper builds upon the Bayesian model in Seshadri et al. 2020c in two respects.
\begin{enumerate}
\item First, we recognise that the area average as a concept lacks theoretical grounding, and attempts to convert a calculated area average to more appropriate averages are not generalisable. Thus, we introduce an approach for computing a mass average. We call this the Bayesian mass average. Although, our goal will be to evaluate \eqref{equ:mass_temp} without fully knowing $T\left(r, \theta \right)$ or indeed $\dot{m}$, the ideas in this paper can be used to calculate \eqref{equ:cap_temp} and \eqref{equ:press_w} as well. 
\item Second, our approach will yield instantaneous mass average estimates, using only available sensor data at a given axial plane. We view this as critical for use in engine measurement and performance testing scenarios. Practically, we offer two distinct scenarios for computing \eqref{equ:mass_temp}, under varying assumptions and data availability.
\end{enumerate}
The remainder of this paper is structured as follows. A cursory summary of the key principles behind Bayesian inference is presented in~\ref{sec:bi}. This sets the stage for the proposed mass averaging formulations that are introduced in~\ref{sec:mflow}, which is followed by numerical examples in~\ref{sec:numex}. 

\section{Bayesian inference}
\label{sec:bi}
This section presents a brief overview of the key ideas underpinning Bayesian inference. For further details, interested readers are encouraged to consult Gelman et al. \cite{gelman2013bayesian} and Kruschke \cite{kruschke2014doing}.

\subsection{Bayes' rule, prediction and likelihood}
Let us assume the existence of a model that yields the spatial distribution of a thermodynamic quantity. Further, let $\alpha$ denote an unobservable vector-valued input to this model. In other words, let $\alpha$ be our model parameters. Given the rather limited information afforded to us, it seems rational to assign $\alpha$ a probability distribution $\mathbb{P}\left(\alpha\right)$ rather than a fixed value. This is called the prior.  

The observations from the few circumferentially- and radially-placed sensors are denoted by the vector $y$, where $y = \left(y_1, \ldots, y_n \right)$ for each of the $n$ sensors. To draw inference regarding $\alpha$ given $y$, one requires a model of the \emph{joint probability distribution} $\mathbb{P}\left(\alpha, y \right)$, which may be expressed as
\begin{equation}
\mathbb{P}\left(\alpha, y \right) = \mathbb{P} \left( \alpha \right) \mathbb{P}\left(y | \alpha \right) = \mathbb{P}\left(y \right) \mathbb{P} \left( \alpha | y \right)
\end{equation}
where the term $\mathbb{P}\left(y | \alpha \right)$ is called the \emph{data distribution}. When interpreting this distribution for a fixed $y$ strictly as a function of $\alpha$, one refers to $\mathbb{P}\left(y | \alpha \right)$ as the \emph{likelihood function} \cite{gelman2013bayesian}. In Bayesian inference, we are interested in ascertaining the conditional probability of the model parameters $\alpha$ given observations $y$
\begin{equation}
\mathbb{P} \left( \alpha | y \right) = \frac{  \mathbb{P} \left( \alpha \right) \mathbb{P}\left(y | \alpha \right) }{\mathbb{P} \left( y \right) },
\label{equ:bayes}
\end{equation}
which is know as Bayes' rule; $\mathbb{P} \left( \alpha | y \right)$ is termed the \emph{posterior density}. Succinctly stated, the objective of Bayesian inference is to develop $\mathbb{P} \left(\alpha, y \right)$ and undergo calculations to determine $\mathbb{P}\left( \alpha | y \right)$ (see pages 6-7 of \cite{gelman2013bayesian}). Note that the denominator in \eqref{equ:bayes} can also be expressed as an integral (also termed marginalization)
\begin{equation}
\begin{split}
\mathbb{P} \left( y \right) & = \int \mathbb{P} \left( y | \alpha \right) \mathbb{P} \left( \alpha \right) d \alpha  \\
& \propto \mathbb{P} \left( y | \alpha \right) \mathbb{P} \left( \alpha \right).
\end{split}
\end{equation}
where we refer to $\mathbb{P} \left(y \right)$ as one of the \emph{marginal distributions} of the joint distribution $\mathbb{P}\left(\alpha, y \right)$.
\subsection{Gaussian process regression}
\label{sec:gp}
Gaussian processes represent a powerful and flexible class of models for predicting spatially- and temporally-varying scalar- and vector-valued functions, where any finite-dimensional marginal distribution of the covariates is a Gaussian distribution \cite{gelman2013bayesian}. A Gaussian process $f$ for some quantity of interest can be defined in terms of its mean $\mu_{f}$ and covariance function $\Sigma_{f}$ \cite{rasmussen2006gaussian}, i.e.,
\begin{equation}
f \left( x \right) \sim \mathcal{N} \left( \mu_{f} \left( x \right) , \Sigma_{f}\left( x, x \right)  \right).
\label{equ:def_gp}
\end{equation}
If we write our \emph{training data} as $\mathcal{D} := \left\{ x_i, y_i   \right\} _{i=1}^{n}$ and \emph{testing data} locations as $x^{\ast} = \left(x^{\ast}_{1}, \ldots, x^{\ast}_{m} \right)$, then the predictive mean and covariance functions of the Gaussian process is given by
\begin{equation}
\begin{split}
\mu_{f}\left( x^{\ast} \right) = K\left( x^{\ast}, x \right) \left( K\left(x, x \right) + \Sigma \right)^{-1} y \\
\Sigma_{f}\left( x^{\ast}, x^{\ast} \right) = K\left( x^{\ast}, x^{\ast} \right) - K\left( x^{\ast}, x \right) \left( K\left(x, x \right) + \Sigma \right) K\left( x, x^{\ast} \right).
\end{split}
\label{equ:gp_kernel}
\end{equation}
Here $\Sigma = \sigma^2 \mI$ is a diagonal matrix of the measurement noise associated with each sensor. We assume a Gaussian noise model, i.e., our observations $y$ are corrupted by a zero-mean noise $\mathcal{N} \left(0, \sigma^2 \right)$. The matrix $K\left( \cdot, \cdot \right)$ in \eqref{equ:gp_kernel} is the \emph{kernel function} $k$ evaluated at the input arguments. This function dictates the covariance between any two points and controls the smoothness of samples from the Gaussian process along with the extent of shrinkage towards the mean \cite{gelman2013bayesian}. Kernel functions are typically parameterized by \emph{hyperparameters} that need to be tuned based on the data. 

The kernel functions used in this paper are based on those in Seshadri et al. 2020c \cite{seshadri2020c}, where the kernel above is written as a product of two kernels
\begin{equation}
K\left(x, x' \right) = K_{r} \left(r, r' \right) \times K_{c} \left( \theta, \theta' \right),
\label{equ:kernel_product}
\end{equation}
where with a slight abuse of notation we imply that $x = \left(r, \theta \right)$. In \eqref{equ:kernel_product} $K_r$ represents the radial kernel, while $K_c$ is the circumferential kernel. Given the inherent differences in the variation of thermodynamic quantities along these two directions, two mathematically distinct kernels are required. The radial kernel is given by a squared exponential 
\begin{equation}
K \left(r,r'\right) = \sigma^2_{f} \; \textrm{exp}\left(-\frac{1}{2l^2} \left(r - r' \right)^2 \right),
\end{equation}
with two \emph{hyperparameters}: a correlation length $l$ and a kernel noise $\sigma_f$. These are assigned the following prior distributions
\begin{equation}
l \sim \mathcal{N}^{+} \left(0, 1 \right) \; \; \; \; \text{and} \; \; \; \; \sigma_f \sim \mathcal{N}^{+} \left(0, 1 \right),
\end{equation}
where the symbol $\mathcal{N}^{+}\left(0, 1 \right)$ indicates a half normal distribution with a mean of zero and a variance of unity. In the circumferential direction, the kernel function is given by
\begin{equation}
K \left( \theta, \theta' \right) = F \left( \theta \right) \Lambda F\left( \theta' \right)^{T},
\label{equ:kernel_fourier}
\end{equation}
where $F$ is a Fourier matrix of the form
\begin{equation}
F\left(\theta \right) = \left( \begin{array}{cccccc}
1 & \textrm{sin}\left(\omega_{1}\theta\right) & \textrm{cos}\left(\omega_{1}\theta\right) & \ldots & \textrm{sin}\left(\omega_{k}\theta\right) & \textrm{cos}\left(\omega_{k}\theta\right)\end{array}\right),
\end{equation}
where $\left( \omega_1,\ldots, \omega_{k} \right)$ represents the $k$ frequencies of the circumferential harmonics. In \eqref{equ:kernel_fourier}, $\Lambda=\text{diag}\left( \lambda_1, \ldots, \lambda_{2k+1} \right)$ is a diagonal matrix of priors assigned to the constant, and the sine and cosine components of each frequency. Following \cite{seshadri2020c}, we set these priors
\begin{equation}
\lambda_j \sim \mathcal{N}^{+} \left(0, 1 \right) \; \; \; \; \text{for} \; \; \;  j=1, \ldots, 2k + 1.
\end{equation}
For ease in notation, we group all our priors into $\mathbb{P}\left(\alpha \right)$, where $\alpha = \left(l, \sigma_f, \lambda_{1}, \ldots, \lambda_{2k+1} \right)$, and thus with a slight abuse in notation we write $\mathbb{P}\left( \alpha \right) = \mathcal{N}^{+} \left(0, \boldsymbol{1}_{2k+3} \right)$, which is a $\left(2k+3\right)$-dimensional standard half normal distribution where each component is independent and has a variance of unity. This sets the stage for a computational approach for estimating the posterior distributions of our hyperparameters.
\subsection{Bayesian inference in Gaussian process regression}
As Gaussian process regression is a strictly non-parametric framework, the model parameters $\alpha$ in \eqref{equ:bayes} cannot be directly substituted. In the Gaussian process model given by $f$ in 
\eqref{equ:def_gp}, the likelihood function is given by $\mathbb{P} \left( y | f, \sigma \right) = \mathcal{N} \left(f, \sigma^2 I \right)$ and the prior at the observed inputs is $\mathbb{P}\left(f | x, \alpha \right) = \mathcal{N} \left(0, K \right)$. The conditional evidence of the Gaussian process can then be computed via
\begin{equation}
\mathbb{P} \left( \mathcal{D} | \sigma, \alpha \right) := \mathbb{P}\left( y | x, \sigma, \alpha \right) = \int \mathbb{P} \left( y | f, \sigma \right) \mathbb{P}\left(f | x, \alpha \right) \mathbb{P}\left(\alpha \right) df \; d\alpha.
\label{equ:integral}
\end{equation}
The expression on the left hand side is the probability of observing the data, conditioned upon the hyperparameter values $\alpha$, and the kernel encoded in $\mathbb{P}\left(f | x, \alpha \right)$. The standard approach for integrating \eqref{equ:integral} is to use Markov chain Monte Carlo (MCMC), from which posterior probability densities of the hyperparameters can be obtained. The challenge with MCMC however is that it may necessitate long running times, particularly as the number of frequencies increase---increasing the number of hyperparameters that need to be tuned. One workaround is to focus on solving a gradient-based optimisation problem for identifying the mode of $\mathbb{P} \left( \mathcal{D} | \sigma, \alpha \right)$ rather than its full distribution \cite{kuss2006gaussian}. This is known as the \emph{maximum aposteriori} (MAP) estimate and is given by
\begin{equation}
\underset{\alpha}{\text{maximize}} \; \; \;  \mathbb{P} \left( y | f, \sigma \right) \mathbb{P}\left(f | x, \alpha \right) \mathbb{P}\left(\alpha \right).
\end{equation}
This optimisation problem is generally non-convex, but automatic differentiation methods can be used to extract gradients that can help guide the optimiser. Further particulars on the assumptions underpinned by MAP and its similarity to maximum likelihood estimation can be found in \cite{rasmussen2006gaussian}. 

Our adoption of MAP in this paper, instead of MCMC, is motivated the need to compute real-time mass average estimates during engine and rig tests. This is driven by the speed of MAP as a strategy compared to MCMC. To clarify, both approaches require prescribing the dominant circumferential modes and thus strictly from an input perspective, they are identical. However, while the MAP estimate will return an optimised single value of $\alpha$, MCMC will return all probable values of $\alpha$.

\section{Bayesian mass average}
\label{sec:mflow}
Let $\xi = \xi \left(r, \theta \right)$ represent our thermodynamic quantity of interest, i.e., temperature, pressure, enthalpy, or entropy. We wish to compute $\xi$'s mass average across the annulus at a fixed axial plane. As before, we assume access to a set of measurements (training data) $\mathcal{D} := \left\{ x_i, y_i   \right\} _{i=1}^{n}$, where $x_i = \left(r_i, \theta_i \right)$ represents the annular location of the measurement and $y_i$ represents the measured thermodynamic quantity $\xi$. For simplicity, we assume that our integration bounds in the radial direction vary from $\left[0,1\right]$ instead of $\left[r_{i,}r_{o}\right]$. This results in the following substitution
\begin{equation}
\bar{\xi}_{\dot{m}}=\frac{\left(r_{o}-r_{i}\right)\int_{0}^{2\pi}\int_{0}^{1}\xi\left(r,\theta\right)\dot{m}\left(r,\theta\right)u\left(r\right)drd\theta}{\left(r_{o}-r_{i}\right)\int_{0}^{2\pi}\int_{0}^{1}\dot{m}\left(r,\theta\right)u\left(r\right)drd\theta}
\end{equation}
where $u\left(r\right)=r\left(r_{o}-r_{i}\right)+r_{i}$. The goal of this section is to compute $\bar{\xi}_{\dot{m}}$ when $\xi\left(r,\theta\right)$ is a Gaussian process
\begin{equation}
\xi \left(r, \theta \right) = \mathcal{N} \left( \mu_{\xi} \left(r, \theta \right), \Sigma_{\xi} \left(r, \theta \right)\right).
\end{equation}

\subsection{Polynomial mass flow rate distribution}
To simplify matters, we adopt the rationale that the mass flow rate distribution is uniform in the circumferential direction. Note that this may not be a safe assumption, and we defer to the reader's judgement on whether this assumption may be valid for their application. In practice, a circumferential variation may be expected, just as is for pressure measurements (see Section 4 in \cite{sae2017}).

If valid, then we can express the mass flow rate distribution as 
\begin{equation}
\bar{\xi}_{\dot{m}}=\frac{\int_{0}^{2\pi}\int_{0}^{1}\xi\left(r,\theta\right)\dot{m}\left(r\right) u\left(r \right) drd\theta}{\int_{0}^{2\pi}\int_{0}^{1}\dot{m}\left(r\right)u\left(r\right) drd\theta}.
\label{equ:mass_average_eq}
\end{equation}
Note that the integral in the numerator in \eqref{equ:mass_average_eq} can be interpreted as a random Gaussian field multiplied by a linear operator, which is also a Gaussian process. This quantity is then integrated resulting in the joint distribution
\begin{equation}
\begin{split}
\left[\begin{array}{c}
y \\
\psi\int \xi \left(z\right)\dot{m}\left(z\right)  u\left(r\right)dz
\end{array}\right]\sim\mathcal{N}\left(\left[\begin{array}{c}
0\\
0
\end{array}\right],\left[\begin{array}{c}
K\left(x,x\right) + \Sigma \\
\psi\int K\left(x,z\right)\dot{m}\left(z\right) u\left(r \right)dz
\end{array}\right.\right.\\
\left.\left.\begin{array}{c}
\psi \int K\left(x,z\right)\dot{m}\left(z\right) u\left(r \right)dz\\
\psi^2 \int\int K\left(z,z\right)\dot{m}^{2}\left(z\right) u^{2}\left(r \right)dzdz
\end{array}\right]\right),
\end{split}
\end{equation}
where for notational brevity we set $z$ as a proxy for $x = \left(r, \theta \right)$ and
\begin{equation}
\psi = \frac{1}{\int_{0}^{2\pi}\int_{0}^{1}\dot{m}\left(r\right)   u\left(r \right) drd\theta},
\label{equ:psi}
\end{equation}
and the definitions of the kernel $K$ hold from before. Symbolic integration packages such as \texttt{Mathematica} and \texttt{Matlab} can be used to analytically evaluate the closed form expressions above from which the mass averaged quantity is defined by
\begin{equation}
\mu_{ \bar{\xi}_{\dot{m}} } = \psi \int K\left(x,z\right)\dot{m}\left(r\right) u\left(r\right)dz  \cdot \left( K \left(x, x \right) + \Sigma \right)^{-1} y
\label{equ:mean}
\end{equation}
and
\begin{equation}
\begin{split}
\sigma^2_{ \bar{\xi}_{\dot{m}} } = \psi^2 \int\int K\left(z,z\right)\dot{m}^{2}\left(r\right)u^{2}\left(r \right)dzdz - \psi \int K\left(x,z\right)\dot{m}\left(r\right) u\left(r\right)dz  \\
\cdot \left( K \left(x, x \right) + \Sigma \right)^{-1} \psi \int K\left(x,z\right)\dot{m}\left(r\right) u\left(r\right)dz,
\end{split}
\label{equ:covar}
\end{equation}
yielding
\begin{equation}
\bar{\xi}_{\dot{m}} \sim \mathcal{N} \left( \mu_{\bar{\xi}_{\dot{m}} } , \sigma^2_{\bar{\xi}_{\dot{m}} }  \right).
\end{equation}
We remark that the Bayesian area average \cite{seshadri2020c} can be obtained simply by setting $\dot{m}\left(r \right)$ in \eqref{equ:psi}, \eqref{equ:mean} and \eqref{equ:covar} to be unity. 

\begin{figure}
\begin{center}
\includegraphics[scale=0.5]{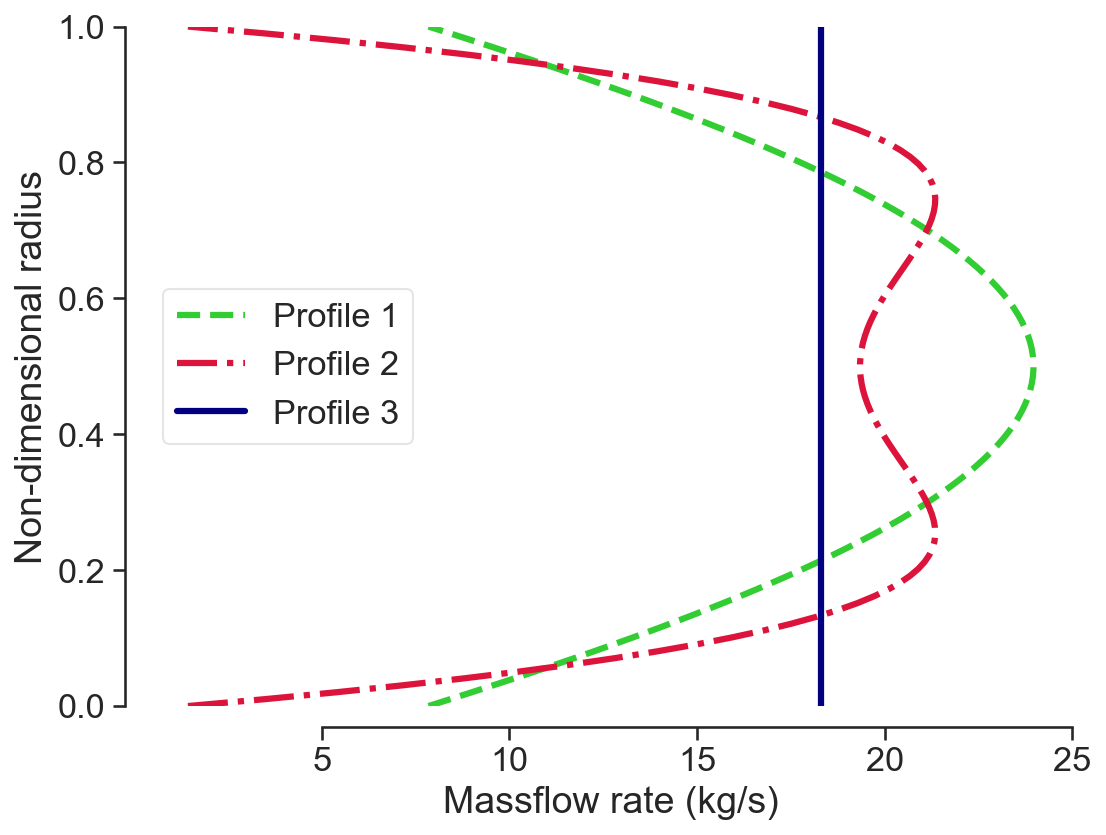}
\caption{Three sample mass flow rate profiles.}
\label{fig:mflow}
\end{center}
\end{figure}

We assume that $\dot{m}$ is (at least) a cubic polynomial that captures the variation in the mass flow rate distribution. If a mass flow rate distribution can be extracted from a representative computational fluid dynamics (CFD) simulation, then the coefficients $\left\{ c_0, c_1, \ldots, c_N   \right\}$ corresponding to the polynomial fit
\begin{equation}
\dot{m} \left( r \right) = c_{0} + c_{1} r + c_{2} r^2 + c_{3} r^3 + \ldots + c_{N} r^{N}
\end{equation}
can be computed using a standard least squares with a Vandermonde matrix. Figure~\ref{fig:mflow} shows a few different sample mass flow rate distributions---each integrate to yield the same mass flow rate area, i.e., they all have the area under the curve. 

It should be noted that one can just as easily derive the integrals in \eqref{equ:mean} and \eqref{equ:covar} using other parametric curves, such as splines
\begin{equation}
\dot{m} \left( r \right) = c_{0} + c_{1} r + c_{2} r^2 + c_{3} r^3 +  \sum_{i=1}^{K} \text{max} \left[ 0, \left(r - s_i \right) \right],
\end{equation}
where $s_i$ corresponds to the spline knots. This does, however, increase the complexity of the expressions that must be integrated. Additionally, note that for certain transonic flows, rakes may register sharp variation in the radial profile, or even a discontinuity. In such a scenario, both a single polynomial or a spline may be inappropriate, and one may have to resort to piecewise polynomials. This would require integrating over the annulus in parts. In other words, rather than have a single integral along the radial direction from $\left[r_{i}, r_{o}\right]$, one may have $S$ intervals of the form $\left[r_{i}, r_{1}\right], \ldots , \left[r_{S}, r_{o}\right]$, where $S-1$ represents the number of radial discontinuities.

\subsection{Random field for mass flow rate}
The Gaussian random field for temperature (or pressure) can be written as $\mathcal{N} \left(\mu_{T} , \Sigma_{T} \right)$ (or $\mathcal{N} \left(\mu_{P} , \Sigma_{P} \right)$). Using the same kernel functions in \cite{seshadri2020c}, or alternate ones, a random field for a mass flow rate can also be obtained. Note that this random field need not be Gaussian. Even if it were Gaussian, as the product of two Gaussian distributions does not yield a Gaussian distribution, there is no closed form analytical solution that one can leverage for the mass average when using a Gaussian random field for the mass flow rate. Thus, in either case a discretised approach must be adopted. 

Let $\left\{ \left(\tilde{\theta}_{1},\tilde{r}_{1}\right),\ldots,\left(\tilde{\theta}_{H},\tilde{r}_{H}\right)\right\} $ be $H$ randomly sampled coordinate locations. Next, let $t_1, \ldots, t_L$ be $L$ independent and identically distributed (IID) random samples from the distribution $\mathcal{N} \left(\mu_{T} , \Sigma_{T} \right)$. Each sample $t_i$ is a vector of length $H$, corresponding to the $H$ spatially randomly distributed coordinates. Let $s_1, \ldots, s_L$ also be $L$ IID random samples from the distribution $\mathcal{P}\left( \mu_{\dot{m}}, \Sigma_{\dot{m}}  \right)$, which need not be Gaussian, where $s_i$ is a vector of length $P$ and is the mass flow rate distribution evaluated at the same coordinates corresponding to $t_i$. The mass average of each sample is given by
\begin{equation}
\bar{T}_{\dot{m}, i}= \frac{ t_i^{T} \text{diag} \left( \tilde{r} \right) s_i  }{s_i^{T} \tilde{r}^{T} },
\label{equ:mass_average_samples}
\end{equation}
where $i=1, \ldots, L$, $\tilde{r} = \left(\tilde{r}_1, \ldots, \tilde{r}_H\right)$ and $\text{diag}\left(\cdot\right)$ is a diagonal matrix of the vector-valued argument. These mass average samples can be used to computationally estimate the mass average distribution; we demonstrate this in the following section in this text.

It is important to note that we have not bounded the mass flow rate values, i.e., they can take on both positive and negative values. The latter may be misleading particularly when the mean mass flow rates are low, but their standard deviations are large, implying that there may be realizations of the random field where there is reverse flow, when in practice that is a numerical modelling artefact. One way to tackle this is to construct a random field for $\text{exp}\left( \dot{m} \right)$ instead of $\dot{m}$, or even modelling $\dot{m}$ with a truncated Gaussian distribution. Other approaches may include the use of constraints on the hyperparameters to ensure that the standard deviations are realistic, or even the introduction of certain \emph{ghost data points}. These points are carefully selected fictitious data that are incorporated with existing \emph{real} measurements to ensure that realisations of the random field adhere to known physical characteristics. These ghost points should typically be placed in regions of the flow field where measurements are not available, and where a constraint is necessary, such as in endwalls.

\section{Numerical examples}
\label{sec:numex}
The two aforementioned proposals for prescribing a mass flow rate distribution for arriving at a Bayesian mass average are illustrated below. The codes used to generate the results in this section were developed in \texttt{python} and use the open-source \texttt{pymc3} package \cite{salvatier2016probabilistic} for Bayesian statistical modelling. 

Both proposals are demonstrated on temperatures from an isolated compressor measurement plane, taken from on-ground engine tests. The rakes in both tests were placed at the same in-passage pitchwise location to avoid capturing the higher frequency blade effects, i.e., although the rakes had different circumferential angles, their location relative to the upstream stators were fixed. The consequence of this placement, as mentioned in \cite{seshadri2020a}, is that the probes only see temperature variations associated with engine modes, i.e., leakage flows, upstream struts, casing ovalisation, thermal asymmetries, ground effects, and other hardware variations. Provided the contribution of the blade effects to the mean term of the Fourier expansion are negligible, then accurate averages can be determined using only information afforded from the engine modes.

\begin{figure}
\begin{center}
\begin{subfigmatrix}{2}
\subfigure[]{\includegraphics[]{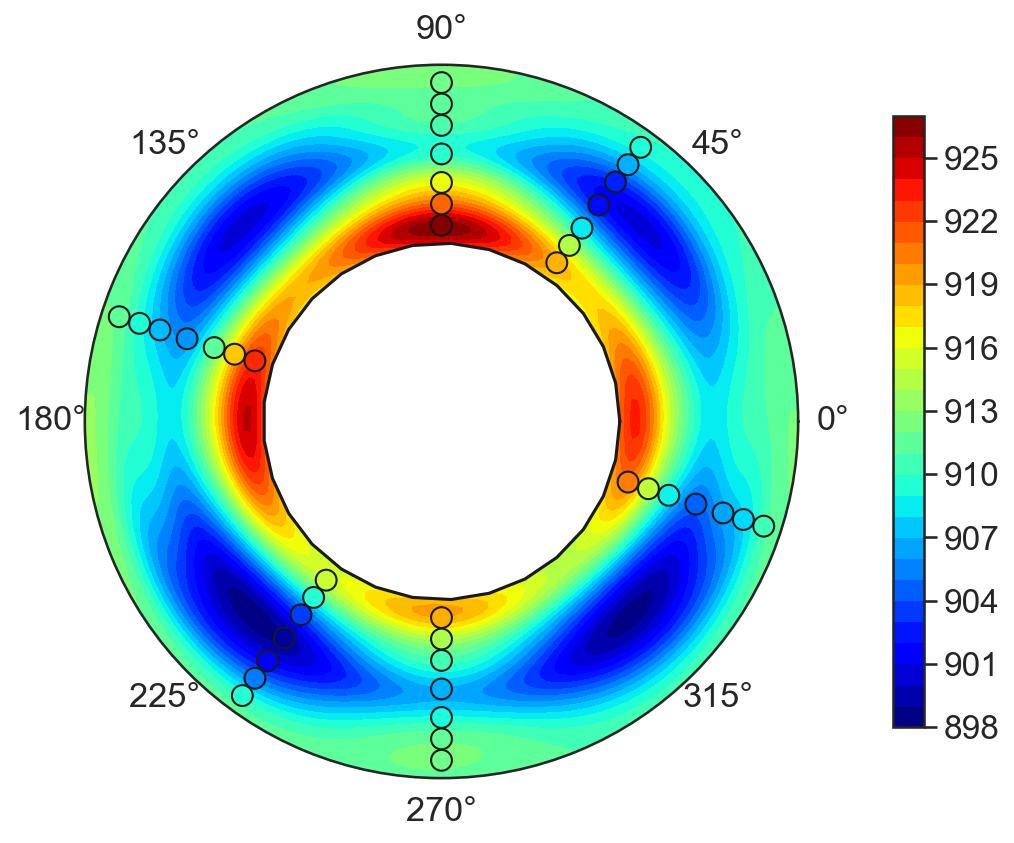}}
\subfigure[]{\includegraphics[]{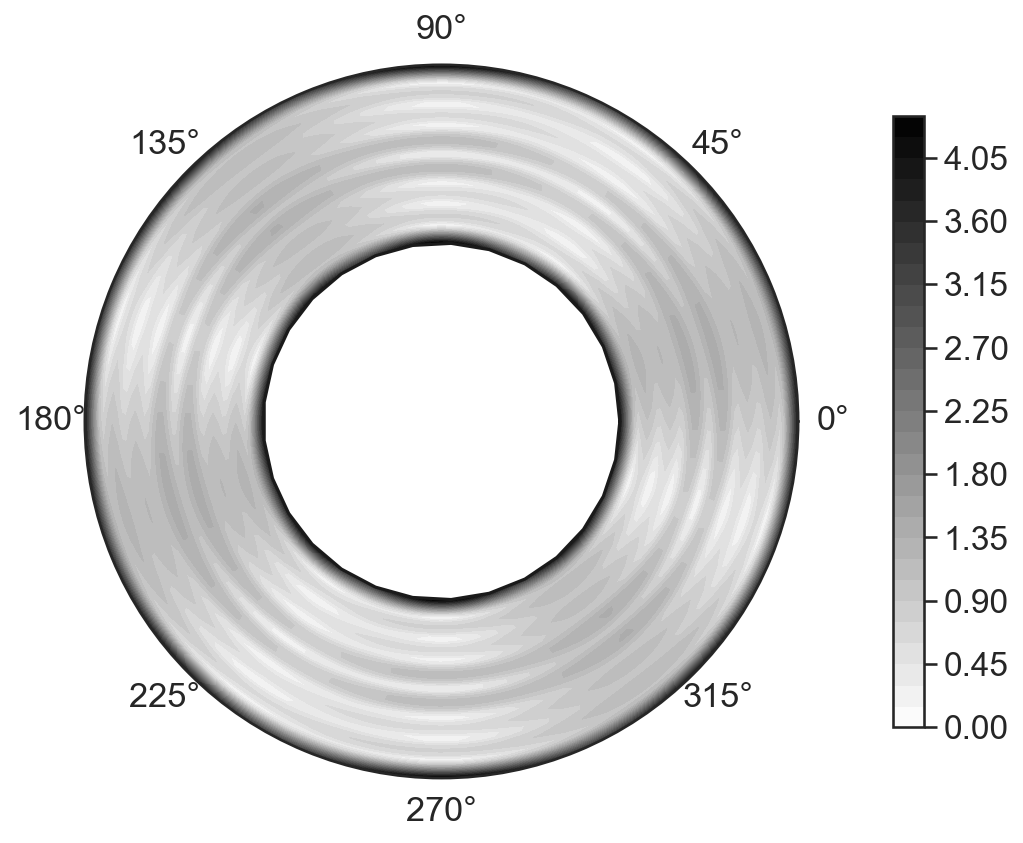}}
\subfigure[]{\includegraphics[]{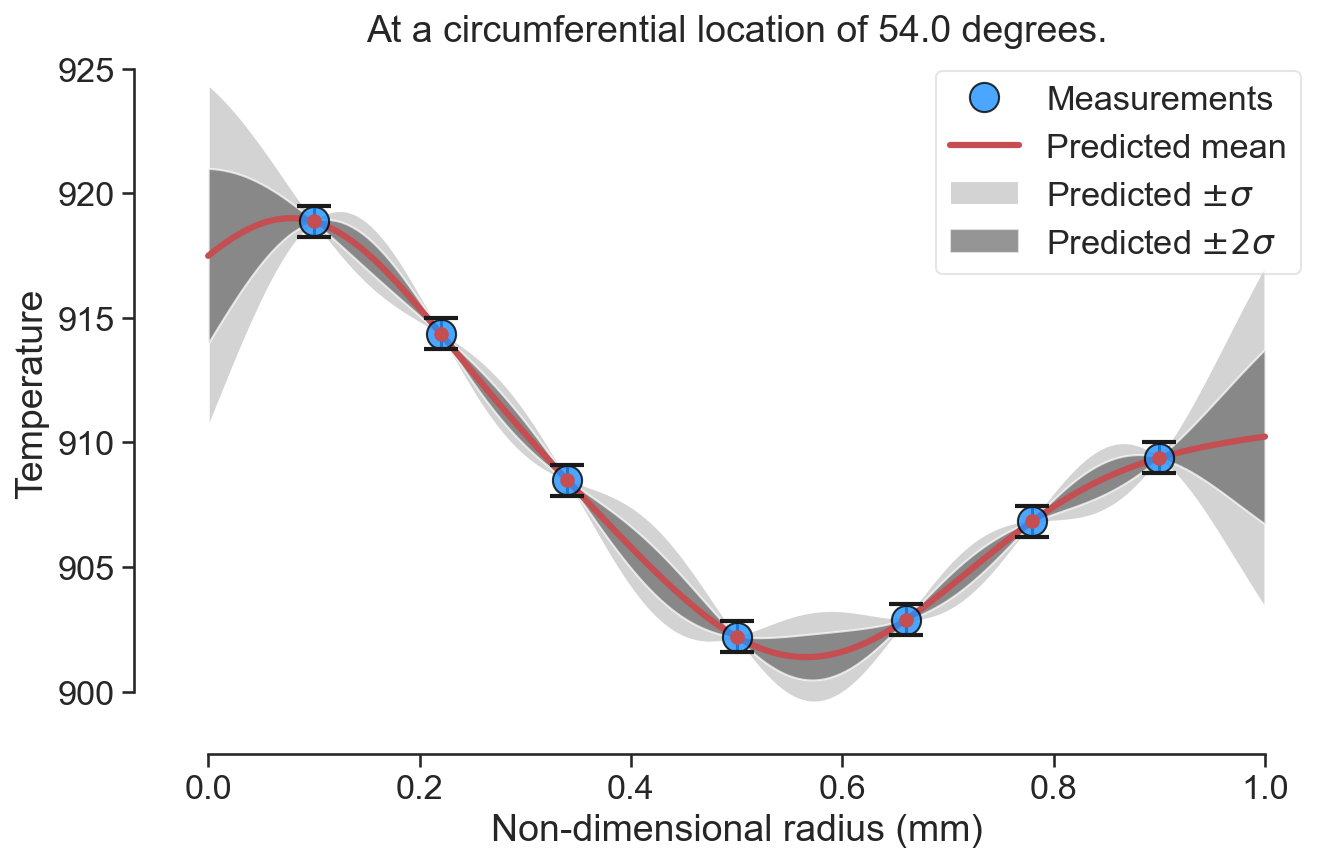}}
\end{subfigmatrix}
\caption{Gaussian process random field distributions for temperature: (a) annular spatial mean; (b) annular spatial standard deviation; (c) Radial slice at a circumferential location of $54^{\circ}$ degrees.}
\label{fig:temp_results}
\end{center}
\end{figure}

\subsection{Polynomial mass flow rate}
Figure \ref{fig:temp_results}(a) and (b) plot the spatial mean and standard deviation in the temperature for the case considered here, assuming frequencies 1, 2 and 4 are present in the pattern---using the priors mentioned in Section \ref{sec:gp}. A radial slice, taken at $54^{\circ}$ degrees is also captured in Figure \ref{fig:temp_results}(c). It is important to note how the Gaussian random field with the prescribed kernels provide an estimate of the uncertainty at the endwalls and between the measurement points that is otherwise difficult to capture.

The three different mass flow rate profiles shown in Figure~\ref{fig:mflow} are applied to the Gaussian random field mentioned above. Profile 1 has a parabolic trajectory, with a peak mass flow rate of 24 kg/sec at mid-span. Profile 2 has a mass flow rate distribution that is more representative of real engine conditions with a particular emphasis at the endwalls where the mass flow rate decreases all the way to zero. Finally, Profile 3 is a uniform mass flow rate distribution in the radial direction. All three mass flow rate profiles are assumed to be uniform in the circumferential direction. Thus, the expectation is that Profile 3 will yield the same mean and variance as the area average, while the other two will likely differ. One particular notion of interest is the difference in the mass average variance relative to the area average variance.

\begin{figure}
\begin{center}
\includegraphics[scale=0.6]{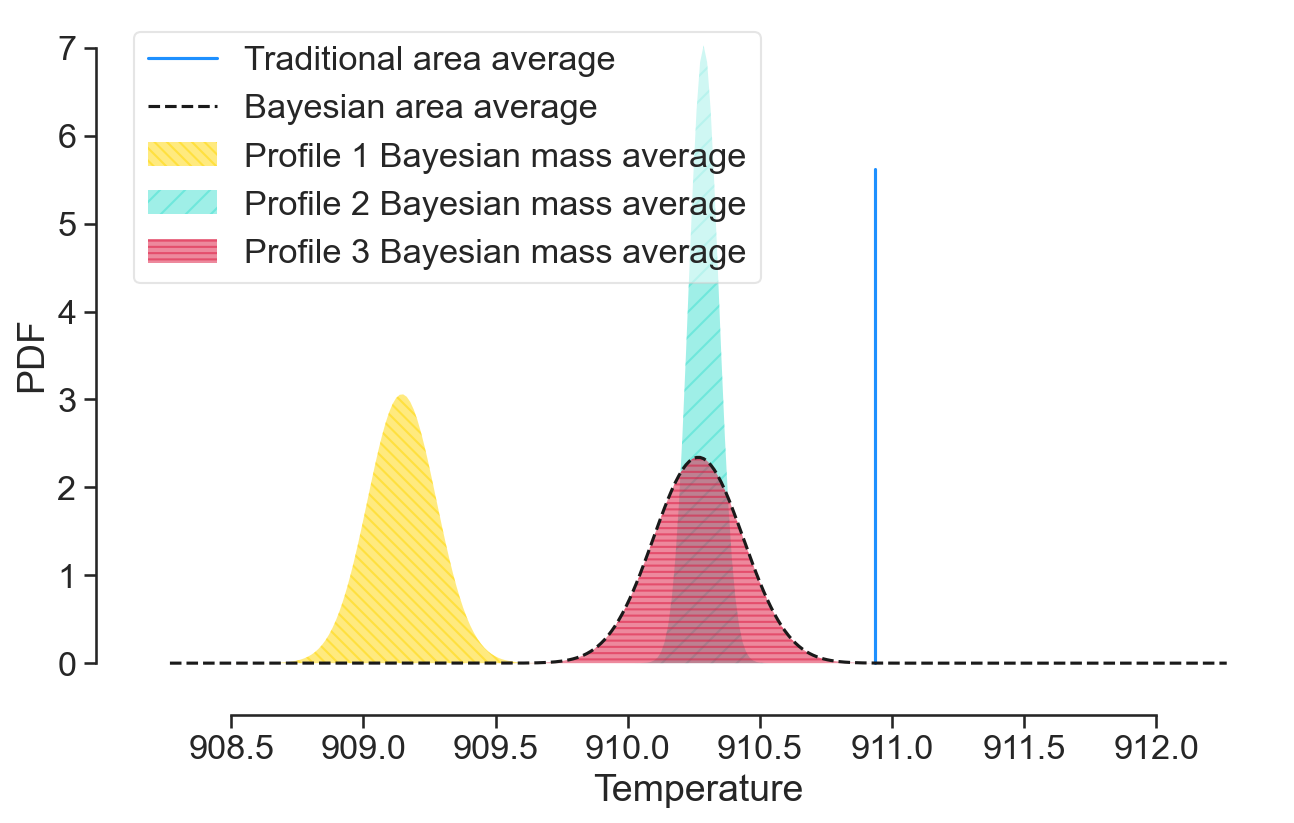}
\caption{Comparison of the area and mass average probability density functions.}
\label{fig:comparison_poly_results}
\end{center}
\end{figure}

The results for the three profiles are shown in Figure~\ref{fig:comparison_poly_results} along with the Bayesian area average. As expected the mean and the variance for the area average are equivalent to that of Profile 3. It is interesting to note that Profiles 1 and 2 lead to distinctly different behaviours---in the former the variance is very slightly reduced compared to the area average, but there is about a 1K difference in the mean. Profile 2, which is more engine-representative, has a very similar mean to the area average, but has a significant reduction in the variance. For completeness, we also include the traditional area average, which as mentioned in \ref{sec:intro}, is computed by weighting each temperature value by the cross-sectional area it covers. 

\subsection{Engine case study}
In this example, we use engine data to inform a velocity distribution that will be used to estimate the mass average temperature. The specific axial plane considered here was fitted with seven circumferential stagnation temperature rakes and three stagnation pressure rakes. As above, the rakes were circumferentially positioned to lie at the same pitch-wise location relative to the upstream stator blades. 

\begin{figure}
\begin{center}
\begin{subfigmatrix}{2}
\subfigure[]{\includegraphics[]{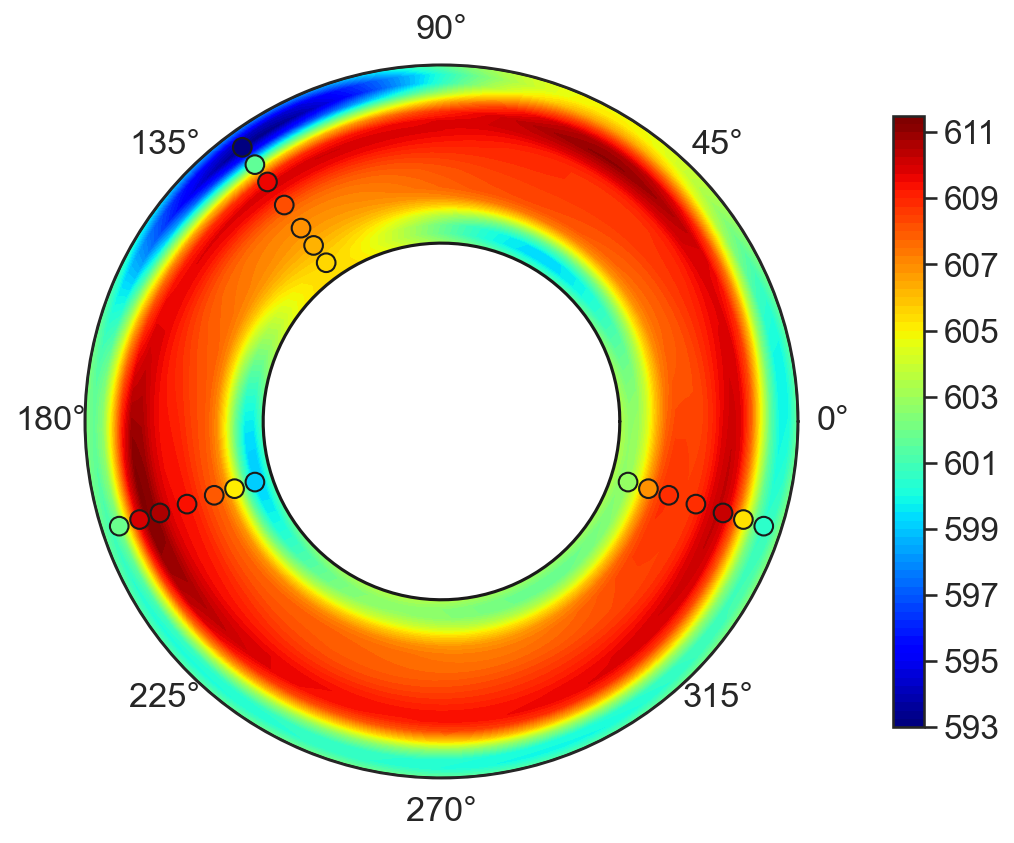}}
\subfigure[]{\includegraphics[]{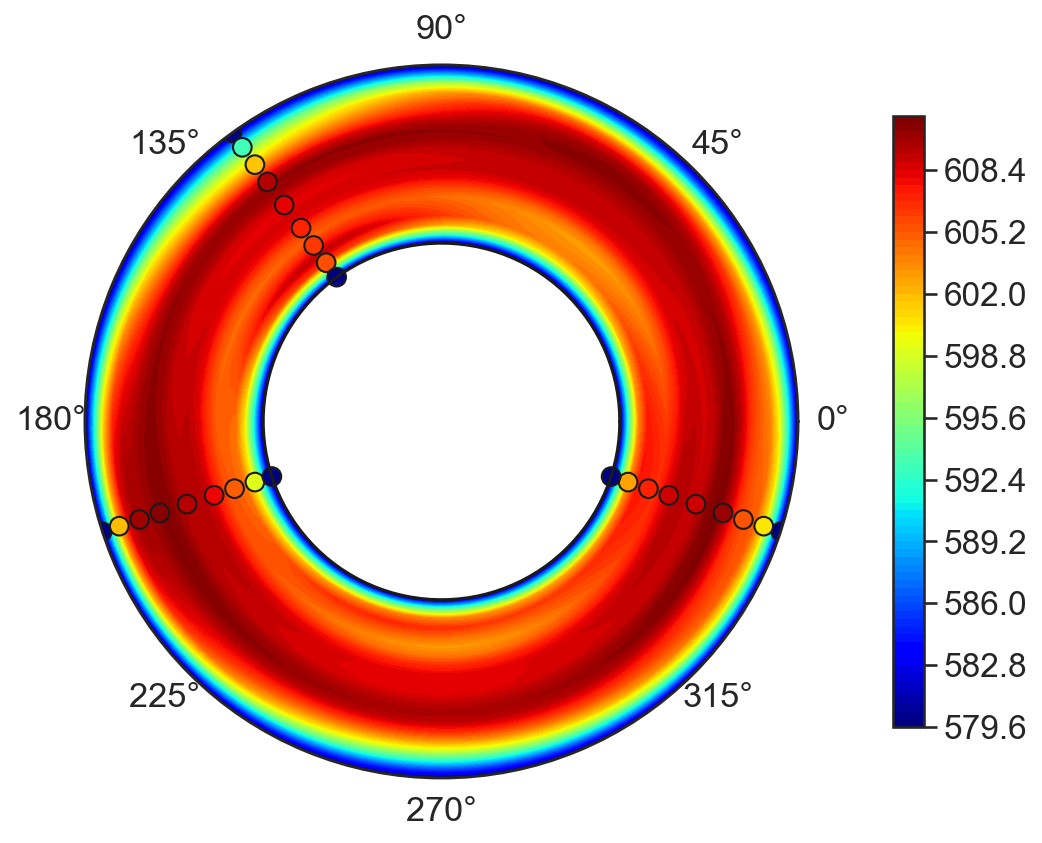}}
\subfigure[]{\includegraphics[]{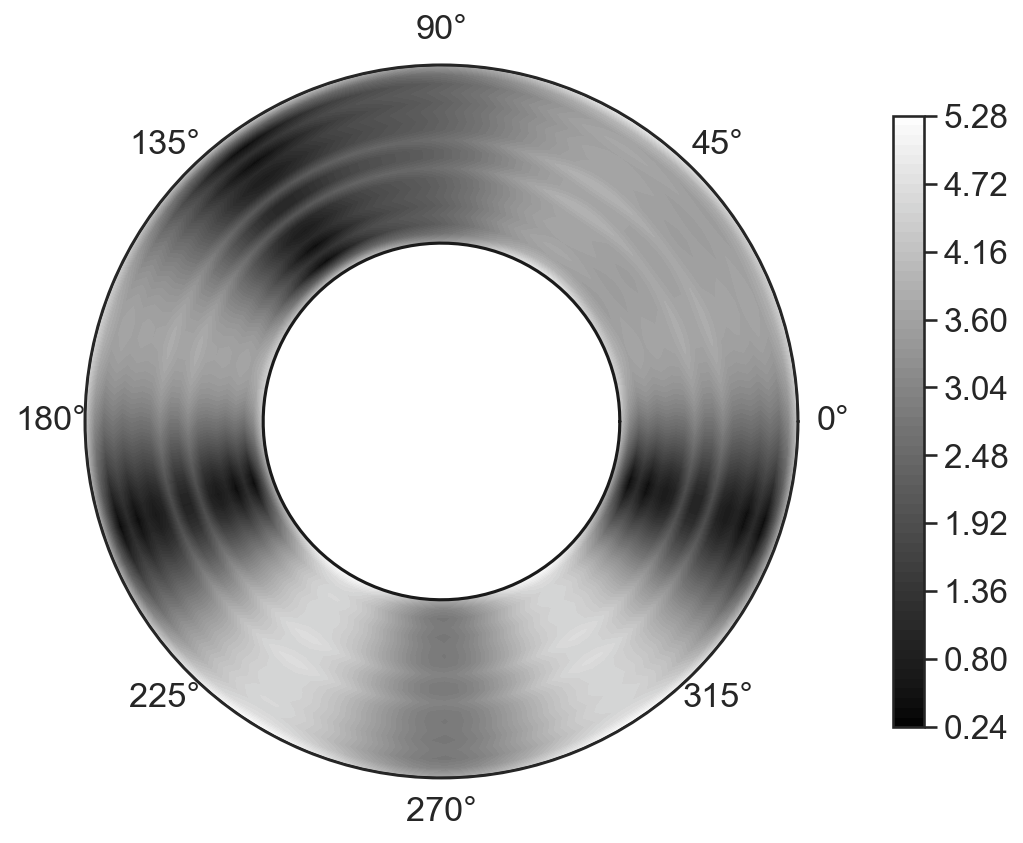}}
\subfigure[]{\includegraphics[]{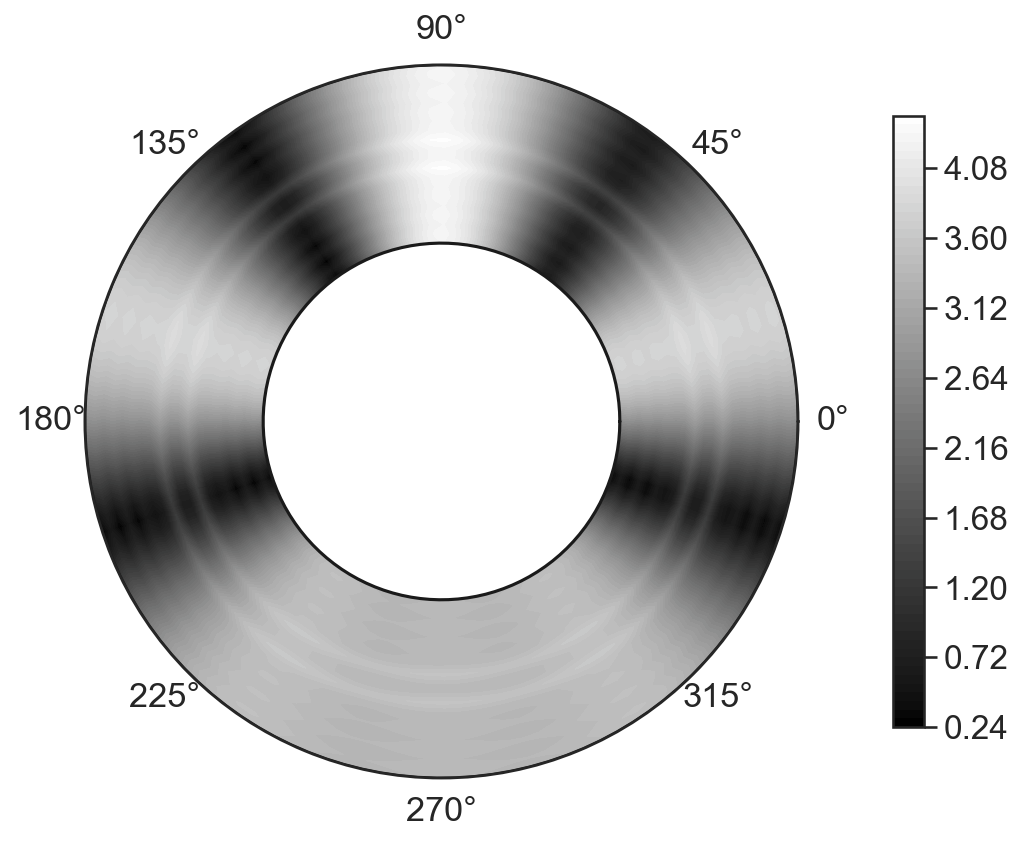}}
\subfigure[]{\includegraphics[]{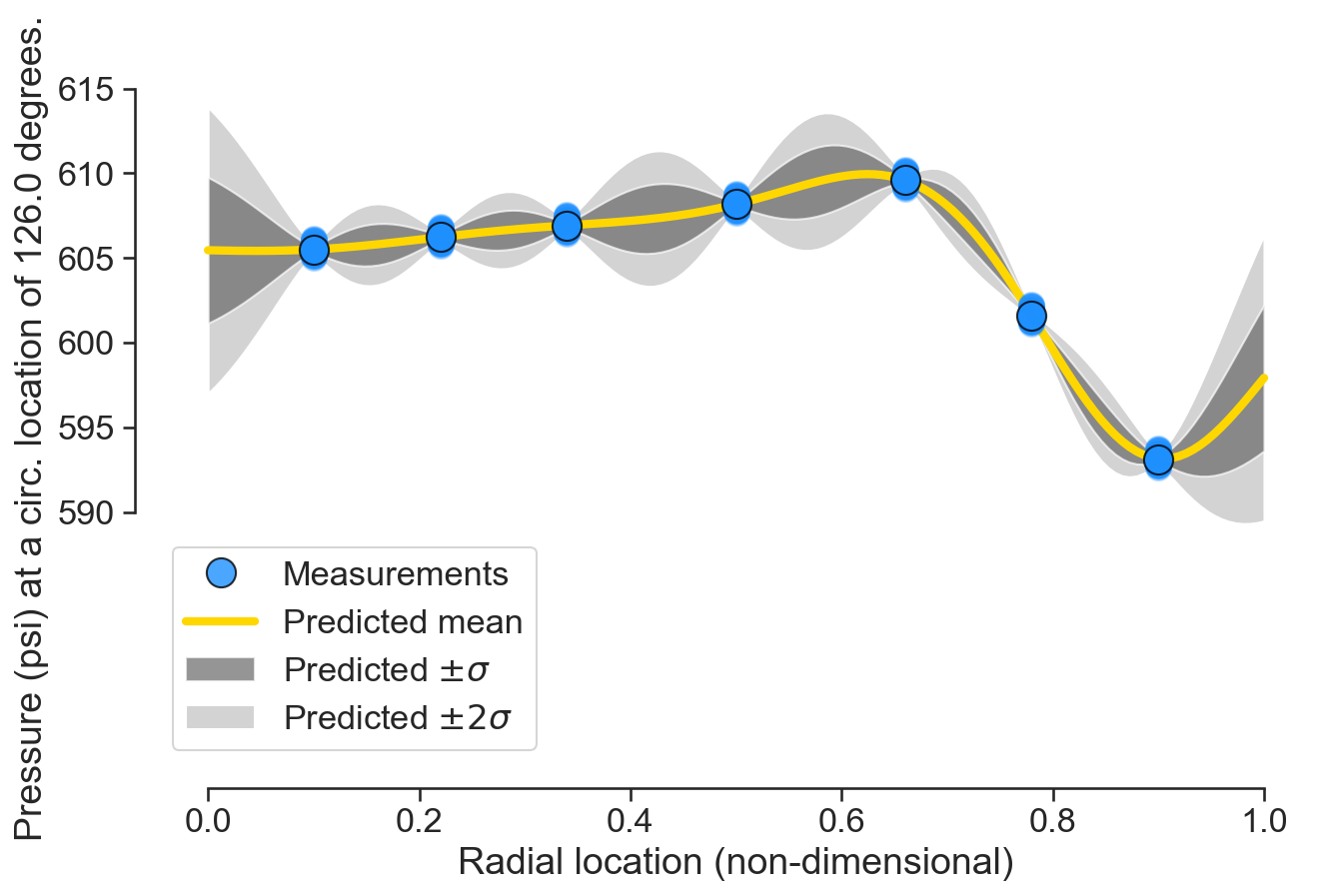}}
\subfigure[]{\includegraphics[]{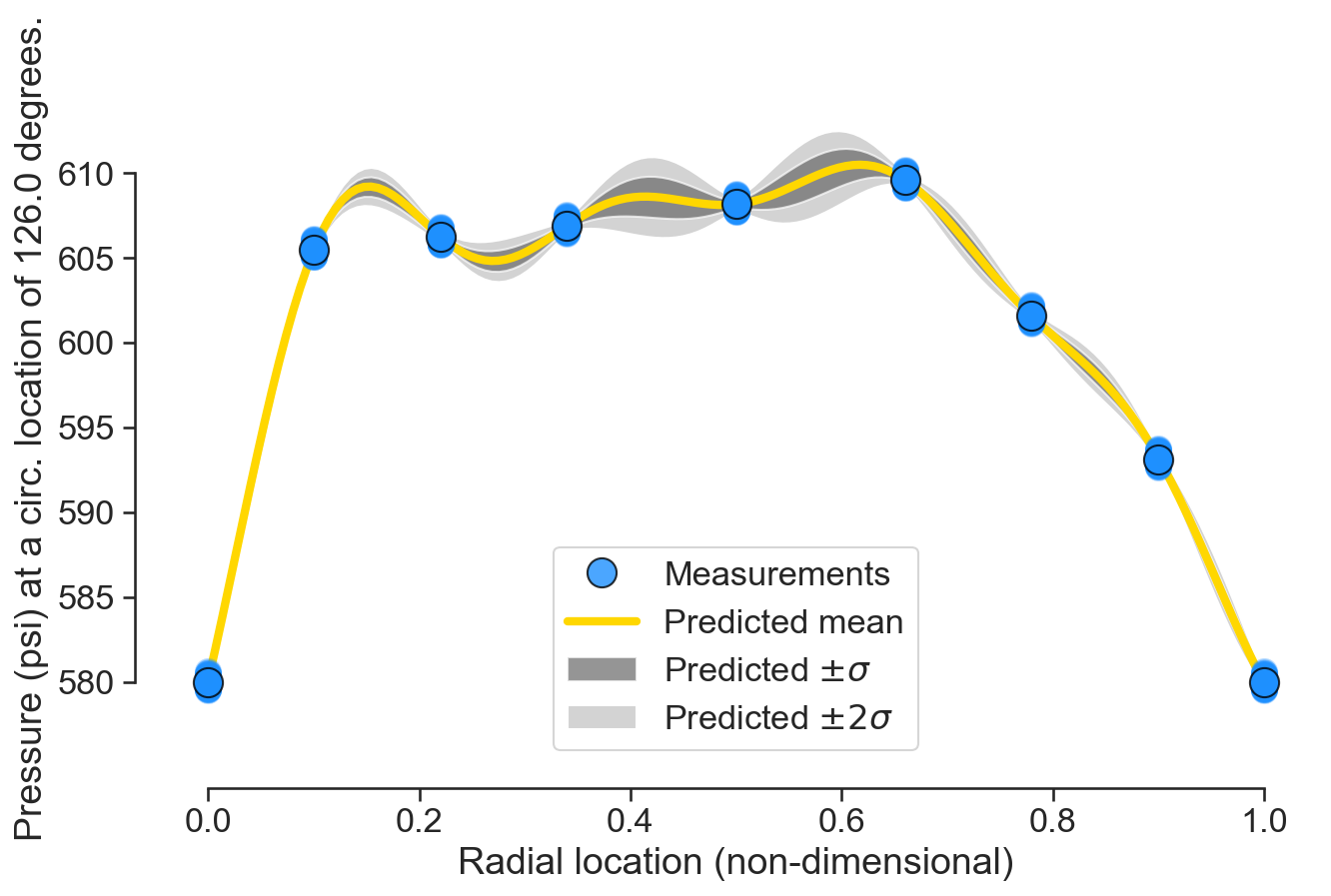}}
\end{subfigmatrix}
\caption{Gaussian random field model for stagnation pressure (in psi) with annular mean (a, b), annular standard deviation (c, d) and radial profiles at $126^{\circ}$ (e, f). The left column is based on real sensor measurements, while the right column two ghost points augmented to the hub and casing for each rake.}
\label{fig:gp_pressure_study}
\end{center}
\end{figure}

We construct a Gaussian random field model to fit the stagnation pressure data, using only modes 1-4 inclusive. Given that there are only three circumferential rakes, the addition of further modes will act to amplify the variance and flatten the mean---an artefact of the bias-variance trade-off. Figures~\ref{fig:gp_pressure_study}(a, c, e) plot the annular mean, annular standard deviation, and the radial distribution of the stagnation pressure corresponding to the rake at $126^{\circ}$. It is clear that there is considerable uncertainty between the hub and the first pressure sensor, and the last pressure sensor and the casing. In fact, one can argue that the pressure profile in these areas is un-physical: one would expect the pressure  to drop to its static value at the endwalls. To remedy this, we append two \emph{ghost point} measurements---one at the hub and one at the casing---to each rake based on our estimate of wall static pressure $\tilde{p}$. In the absence of a known wall static pressure, one can derive the velocity from the stagnation pressure and identify the value which brings the velocity close to zero (see forthcoming paragraph) at the endwalls. When this augmented data is fed into the Gaussian random field model, a more realistic stagnation pressure profile is recovered; see Figures~\ref{fig:gp_pressure_study}(b, d, f). Note that one may choose to replace ghost points above with an appropriately matched turbulent velocity profile. In such a scenario, practitioners may opt for additional ghost points (far greater than two) for best capturing the velocity profile. 

Assuming that the density is constant at this measurement plane, we can estimate the velocity distribution $v \left(r,\theta \right)$ from the isentropic flow equations
\begin{equation}
\frac{P\left(r, \theta \right)}{p\left(r, \theta \right)} = \left( 1 + \frac{\gamma - 1}{2}M\left(r, \theta \right)^2 \right)^{\frac{\gamma}{\gamma - 1}},
\label{equ:isentropic}
\end{equation}
where $p \left(r, \theta \right)$ is the static pressure and $M \left(r, \theta \right)$ is the Mach number. Expressing the latter as a function of the local static temperature $t\left(r, \theta \right)$, we have $M\left(r, \theta \right) = v \left(r, \theta \right) / \sqrt{\gamma \textrm{R} t\left(r, \theta \right)}$. Plugging this into \eqref{equ:isentropic} and re-arranging yields
\begin{equation}
v \left(r, \theta \right) = \sqrt{ \frac{2 \gamma \textrm{R} \; t\left(r, \theta \right)}{\gamma - 1} \left[  \left( \frac{ P\left(r, \theta \right) }{p\left(r, \theta \right)}  \right)^{\frac{\gamma - 1}{\gamma}} - 1\right] }.
\label{equ:vel_first}
\end{equation}
As no measurement information regarding the static temperature $t\left(r, \theta \right)$ is provided, the isentropic flow relations can be used to express this as a function of the stagnation temperature
\begin{equation}
t\left(r, \theta \right) = T \left(r, \theta \right) \left(  1 + \frac{\gamma - 1}{2} M^2 \left(r, \theta \right) \right) =  T \left(r, \theta \right) \left( \frac{ P\left(r, \theta \right) }{p\left(r, \theta \right)}  \right)^{-\frac{\gamma - 1}{\gamma}}. 
\label{equ:static_temp}
\end{equation}
Substituting \eqref{equ:static_temp} into \eqref{equ:vel_first} yields an expression for the velocity field
\begin{equation}
v \left(r, \theta \right) = \sqrt{ \frac{2 \gamma \textrm{R}}{\gamma - 1} \; T\left(r, \theta \right) \left[ 1 -  \left( \frac{ P\left(r, \theta \right) }{p\left(r, \theta \right)}  \right)^{-\frac{\gamma - 1}{\gamma}} \right] }.
\label{equ:vel_first}
\end{equation}
The only unknown that remains is the static pressure $p\left(r, \theta \right)$. For the measurement plane used here, the flow does not have a strong radial or tangential component and is close to axial. Thus, the static pressure field can be assumed constant across the annulus $p\left(r, \theta \right) = \tilde{p}$, permitting us to arrive at the velocity distribution. It should be noted that in scenarios where the above assumption cannot be made, supplemental static pressure measurement data will be required.

\begin{figure}
\begin{center}
\begin{subfigmatrix}{2}
\subfigure[]{\includegraphics[]{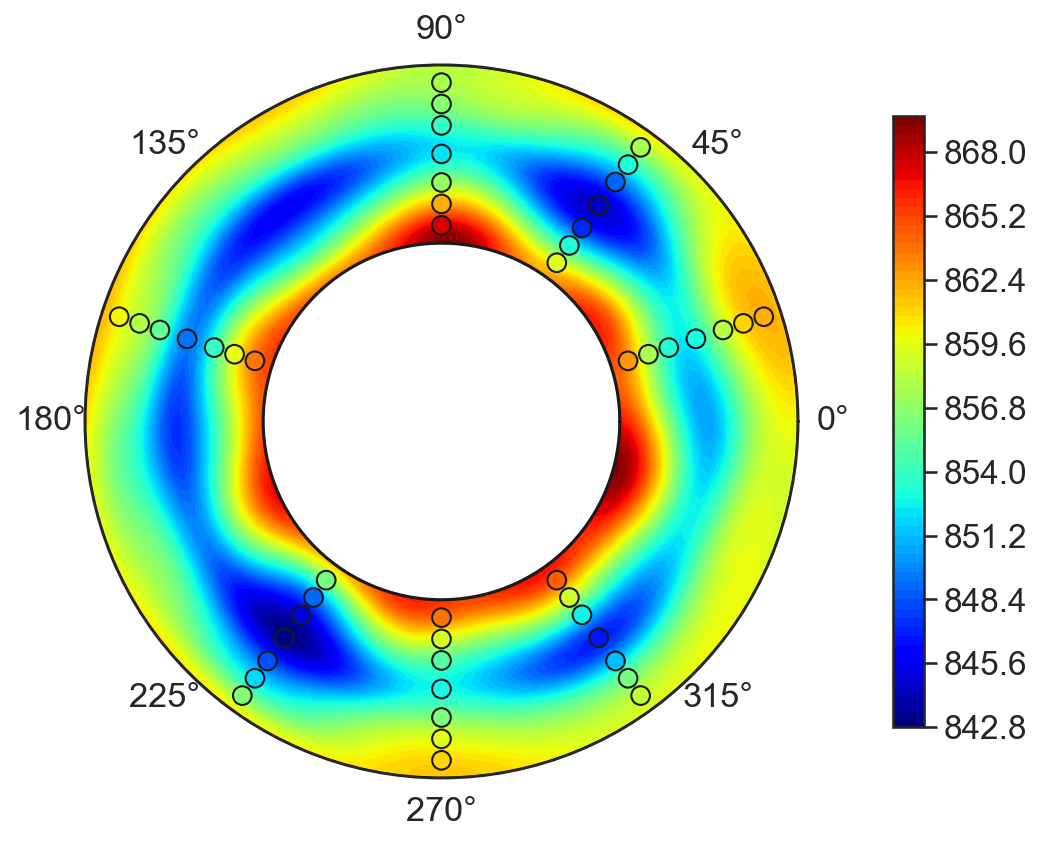}}
\subfigure[]{\includegraphics[]{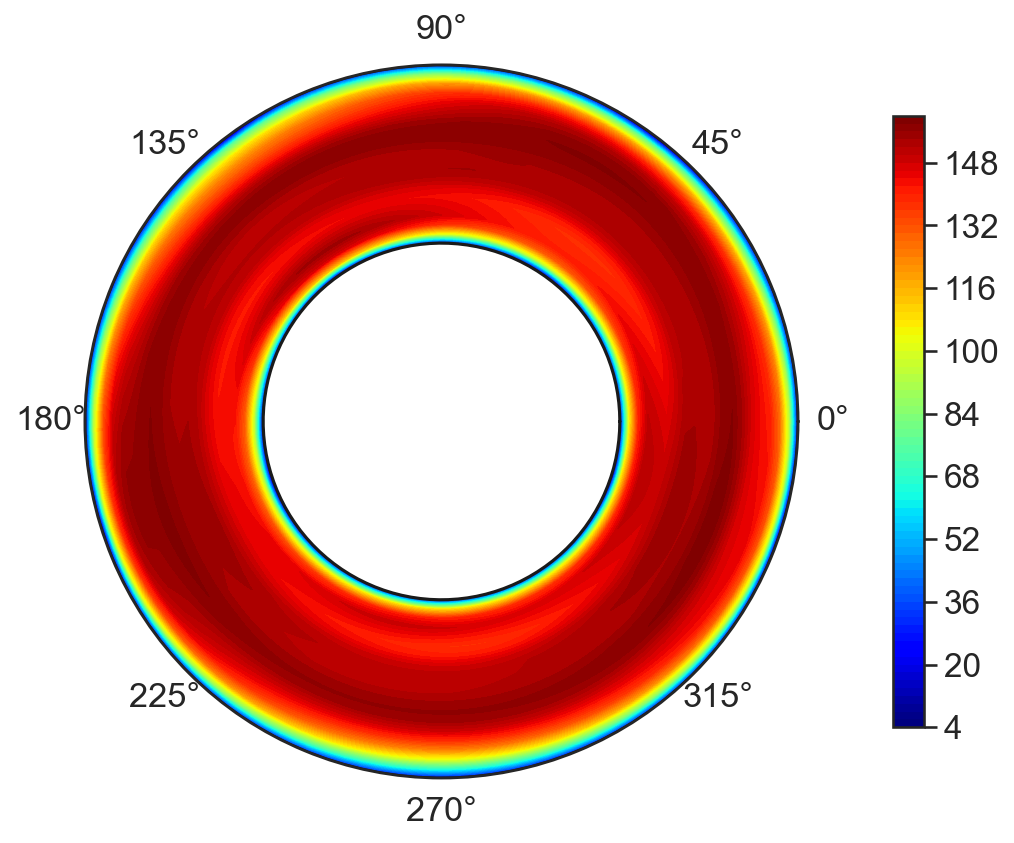}}
\subfigure[]{\includegraphics[]{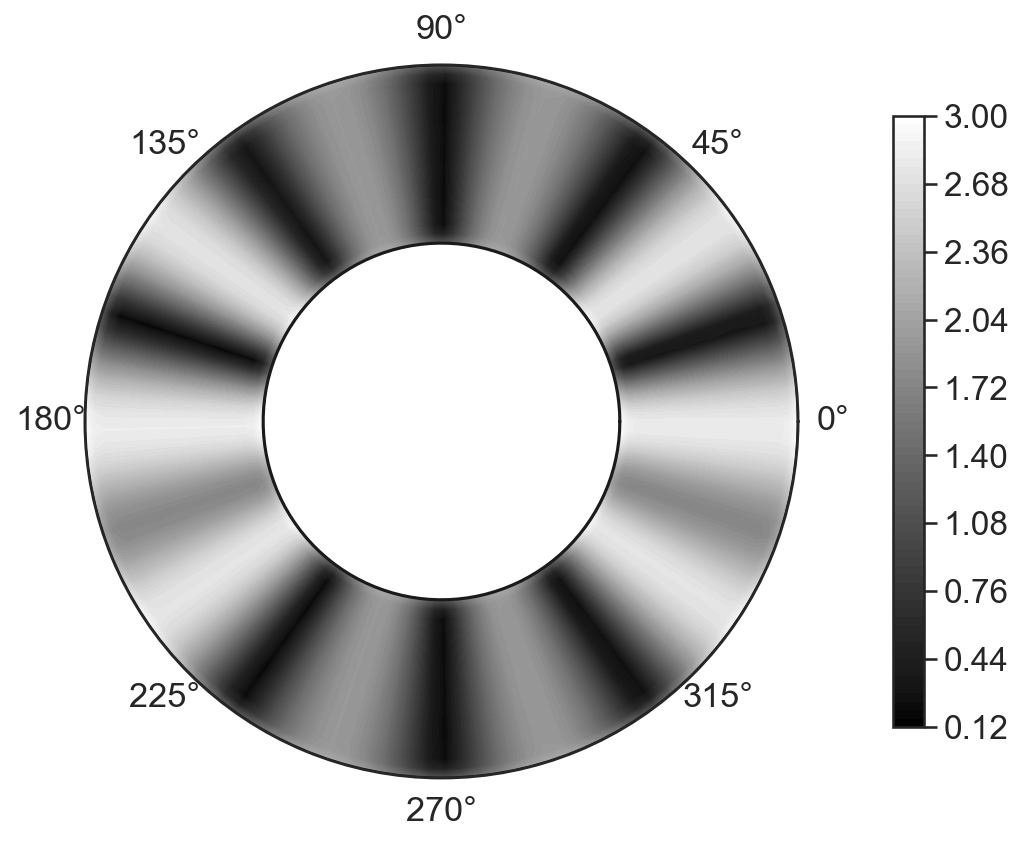}}
\subfigure[]{\includegraphics[]{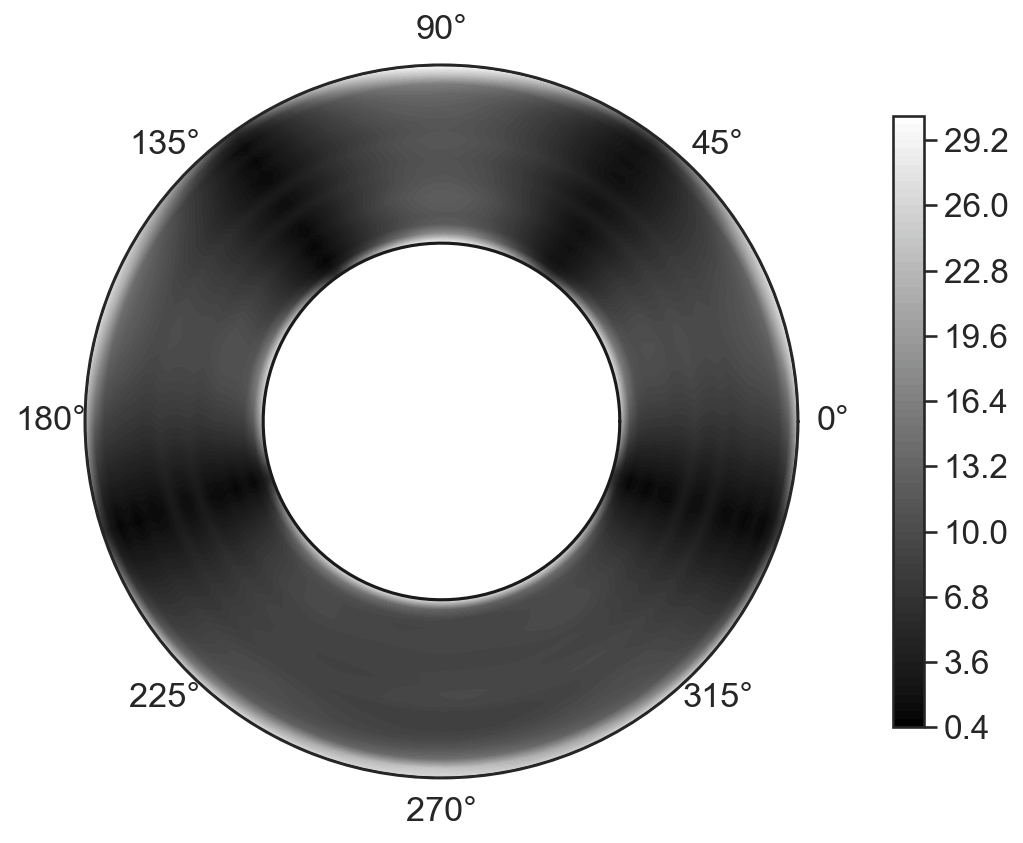}}
\subfigure[]{\includegraphics[]{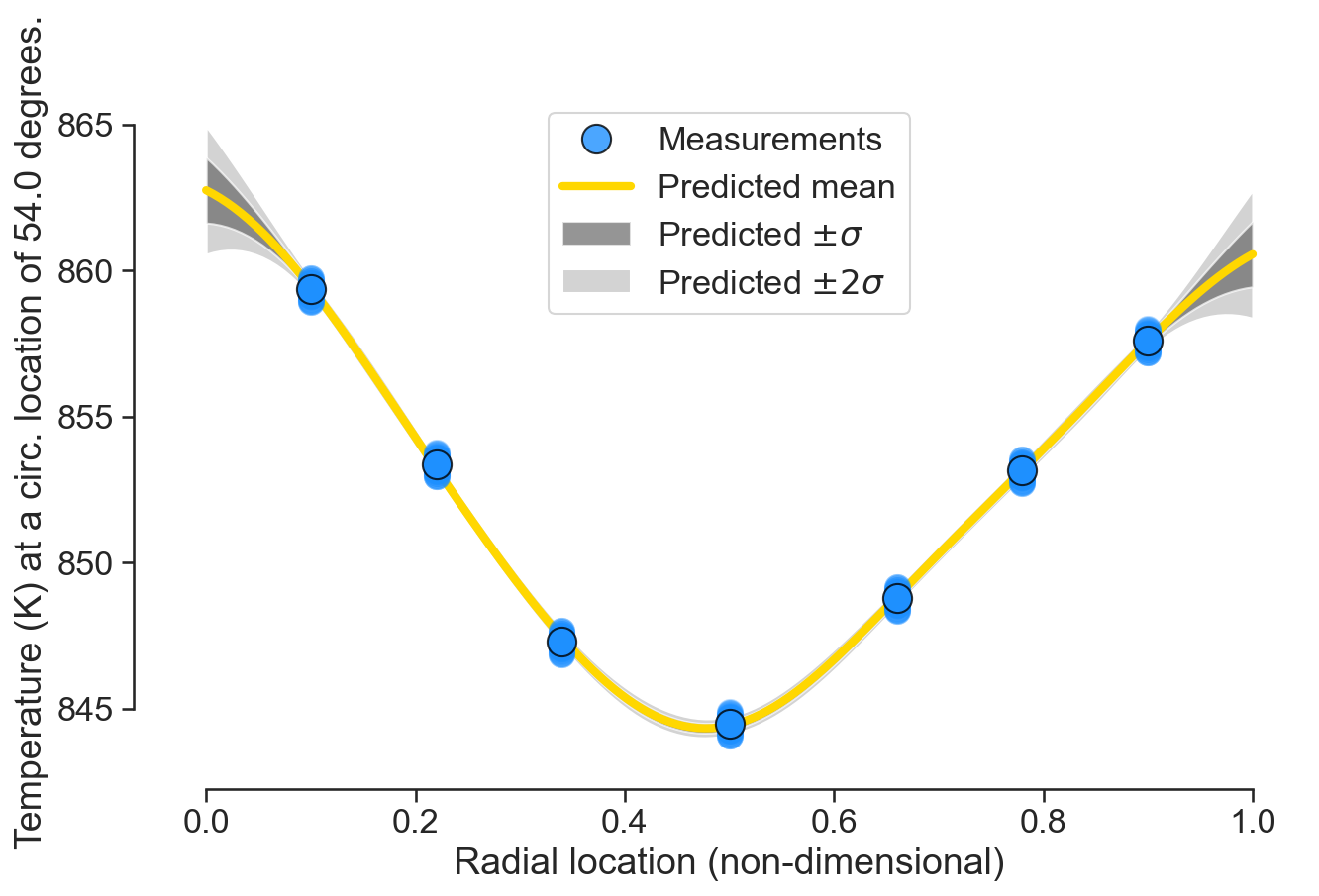}}
\subfigure[]{\includegraphics[]{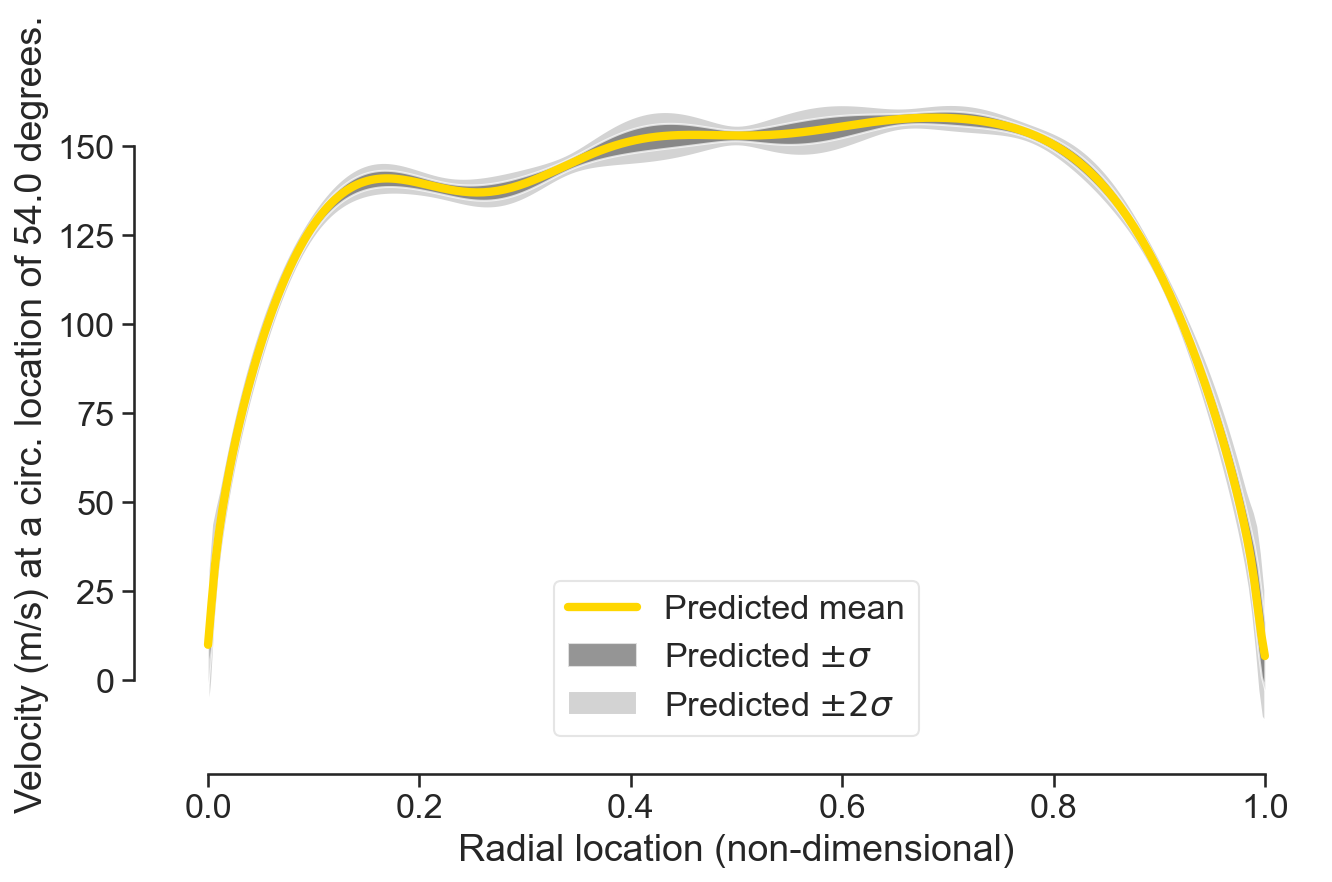}}
\end{subfigmatrix}
\caption{The left column shows the Gaussian random field model for stagnation temperature (in K) with (a) annular mean; (c) annular standard deviation, and (e) radial profile at $54^{\circ}$. The right column shows the (non-Gaussian) random field for velocity with (a) annular mean; (b) annular standard deviation, and (c) radial profile at $54^{\circ}$.}
\label{fig:gp_temp_vel_study}
\end{center}
\end{figure}

A second Gaussian random field is constructed for the stagnation temperature. As there are more temperature rakes available, but yet an uncertainty into the precise circumferential modes, we set the modes to be 1-8 inclusive. In application environments were instant results are not required, a MCMC-based sparse prior formulation would be better suited compared to the used MAP for posterior inference. Using the MAP, we effectively obtain
\begin{align}
\begin{split}
T\left(r, \theta \right) & \sim \mathcal{N}\left( \mu_{T}\left(r, \theta \right), \Sigma_{T}\left(r, \theta, r', \theta' \right) \right), \; \; \; \textrm{and} \\
P\left(r, \theta \right) & \sim \mathcal{N}\left( \mu_{P}\left(r, \theta \right), \Sigma_{P} \left(r, \theta, r', \theta' \right) \right),
\end{split}
\label{equ:gp_temp_press}
\end{align} 
where $\mu_{T}\left(r, \theta \right)$ and $\textrm{diag}[\Sigma_{T}\left(r, \theta, r', \theta' \right)]$ are captured in Figures~\ref{fig:gp_temp_vel_study}(a) and (b) while $\mu_{P}\left(r, \theta \right)$ and $\text{diag}[\Sigma_{P}\left(r, \theta, r', \theta' \right)]$ have already been shown in Figures~\ref{fig:gp_pressure_study}(d) and (e). It should be clear that one simply cannot \emph{plug} the stagnation pressure and temperature values in \eqref{equ:vel_first} with \eqref{equ:gp_temp_press}; although both temperature and pressure are Gaussian processes, velocity is not. Thus, to generate the spatial mean and covariance for velocity, for each $\left(r, \theta \right)$ coordinate, we need to generate multiple samples from the distributions in \eqref{equ:gp_temp_press} and propagate them through the velocity expression. Computing the spatial mean and covariance from those samples yields the velocity profiles shown in Figures~\ref{fig:gp_temp_vel_study}(d-f). As mentioned before, the addition of the \emph{ghost points} to the stagnation pressure data insures that the velocity near the endwalls is close to zero. Once the velocity is known, we evaluate \eqref{equ:mass_average_samples} to estimate the mass average. 

\begin{figure}
\begin{center}
\includegraphics[scale=0.55]{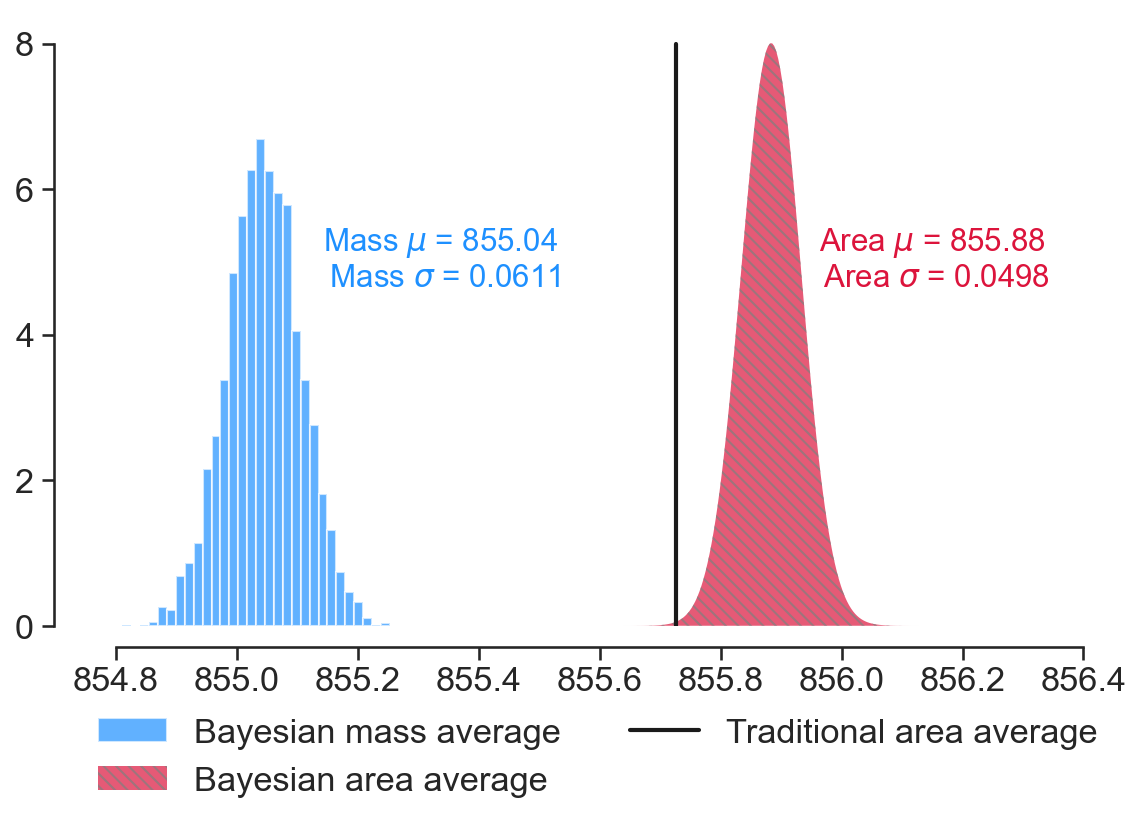}
\caption{Comparison of the Bayesian area and mass average in temperature for the engine case study.}
\label{fig:comparison_results}
\end{center}
\end{figure}

The computed Bayesian mass average is shown in Figure~\ref{fig:comparison_results}; we report a difference of nearly 1K between the Bayesian area and mass average mean owing to the lower temperature values reported at the mid-span regions in Figure~\ref{fig:gp_temp_vel_study}(a). Additionally, we note that the Bayesian mass average is relatively insensitive to large uncertainties in the velocity distribution. The traditional area average (weighted measurements) is also shown for comparison. In this specific case, the difference between the traditional area average and the Bayesian area average is clearly small, however, for the former this is strictly a function of rake placement and thus statements on its accuracy should be made cautiously.

\section{Conclusions}
In this paper, two new methods for more rigorous 
calculations of the mass average are proposed.
\begin{enumerate}
\item The first methodology provides a closed-form analytical formula for the mass average, provided the mass flow distribution is uniform in the circumferential direction, and can be described by a polynomial in the radial direction. 
\item The second methodology does not yield a closed-form analytical formula for the mass average. However, moments of the mass average can be easily determined via sampling. This methodology is tailored for scenarios where: (i) the spatial density can be assumed constant, and (ii) the velocity distribution can be inferred from other measurements, e.g., stagnation pressure.
\end{enumerate}
The results of this paper demonstrate how 1D stagnation quantities calculated via the Bayesian mass average are different from Bayesian area averages and traditional area averages. It facilitates an estimate of the mass average that is rigorous even in the presence of sufficient uncertainty.

\section*{Acknowledgments}
The authors are grateful to Ra\'{u}l V\'{a}zquez-D\'{i}az (Rolls-Royce). The work was part funded by the Fan and Nacelle Future Aerodynamic Research (FANFARE) project under grant number 113286, which receives UK national funding through the Aerospace Technology Institute (ATI) and Innovate UK, together with work funded by Rolls-Royce plc. The authors are grateful to Rolls-Royce plc for permission to publish this paper.

\bibliography{references}
\bibliographystyle{asmems4}

\end{document}